\documentstyle[prl,aps,twocolumn,epsf]{revtex}

\def\eq{\begin{equation}}
\def\eeq{\end{equation}}
\def\eqa{\begin{eqnarray}}
\def\eeqa{\end{eqnarray}}
\def\Scl{{\cal L}}

\def\pref#1{(\ref{#1})}

\begin{document}
\twocolumn[\hsize\textwidth\columnwidth\hsize\csname@twocolumnfalse%
\endcsname
\rightline{McGill-01/08, DIAS-STP-01-08}
\title{Duality and Non-linear Response for Quantum Hall Systems}
\author{C.P.~Burgess$^{1,2}$ and Brian P. Dolan$^{3}$\\
\vspace{3mm}
{\small
$^{1}$ Institute for Advanced Study, Princeton, NJ, 08540.}\\
{\small
$^{2}$ Physics Department, McGill University,
3600 University Street, Montr\'eal, Qu\'ebec, Canada H3A 2T8.}\\
{\small
$^{3}$ Dept. of Mathematical Physics, 
National University of Ireland, Maynooth, Republic of Ireland.}\\
\vspace{3mm}
{\small e-mail: cliff@physics.mcgill.ca, bdolan@thphys.may.ie}}

\date{March 2001}

\maketitle
\begin{abstract}
We derive the implications of particle-vortex duality for the
electromagnetic response of Quantum Hall systems beyond the
linear-response regime. This provides a first theoretical
explanation of the remarkable duality which has been observed in
the nonlinear regime for the electromagnetic response of Quantum Hall 
systems. 
\end{abstract}
\bigskip
\hfill{PACS nos: 73.40.Hm, 05.30.Fk, 02.20.-a} 
\bigskip
]

\section{Introduction}
There is now a good understanding of the physical processes
underlying the quantum Hall effect, at least at the Hall plateau
where the Laughlin wave-functions and the Jain hierarchy give
accurate and insightful descriptions of the observed phenomena.
However the transition between plateaux as external quantities (like
magnetic field) are varied is less well understood theoretically,
despite there now being a wealth of experimental data on these crossovers.

A milestone in this understanding was the observation 
that many features of the critical points of these transitions 
were `superuniversal' \cite{SUniv} (see, however, \cite{Caveat}),
and the subsequent interpretation of this in terms of an underlying
symmetry \cite{KLZ,LutkenRoss}. Unfortunately, the resulting 
understanding of this symmetry in terms of microscopic 
physics \cite{KLZ} -- the Law of Corresponding States -- 
suffers from two related drawbacks. First, it relies on what are
ultimately uncontrolled (mean-field) approximations when making 
contact with Quantum Hall observables. Second, the derivation 
suggests that the domain of validity of the symmetries is more restricted 
than appears to be seen in experiments, being apparently restricted to
the domain of linear response and to the immediate vicinity of the 
critical points of the transitions between plateaux. 

In an earlier paper \cite{PVD} we argued that many of the consequences of
Kivelson Lee and Zhang's analysis could be understood away from the critical
points and without making the mean-field approximation. In a nutshell, 
this was done by deriving them from an effective low-energy theory 
consisting of electromagnetically interacting quasiparticles or vortices. 
In two space dimensions the interactions of particles and vortices are 
described by similar lagrangians, and it is the symmetries which follow 
from this similarity which underlie the success of the Law of 
Corresponding States. These successes may therefore be seen to follow
as predictions for {\it any} system for which the low energy electromagnetic
response can be shown to be well described by the effective
theory we propose. 

Here we extend the discussion of \cite{PVD} 
to applications which are beyond the approximation of linear response. 
In particular we shall derive duality relations for the electromagnetic
response which apply even in the non-linear regime. In so doing we  
provide the first theoretical explanation of the remarkable symmetry under 
interchange of current and (longitudinal) voltage which has been measured
near the critical point of the transition between Laughlin plateaux and 
the Hall Insulator \cite{nonlinear}. 

We begin, in the next section, by reviewing the main features of the
description of the low-energy electromagnetic response given in 
ref.~\cite{PVD}. This is followed -- in section III -- by the 
extension of this description beyond the regime of linear 
response.

\section{The Effective Theory}

Our starting point is the recognition that the energies associated with electromagnetic response experiments are much lower than the typical
microscopic electronic energies. For instance the activation energy as
measured by the temperature dependence of the Ohmic resistivity is of 
order $E_r \sim 0.1 {\rm K} \sim 10 \mu$eV, as compared to the 
underlying Coulomb and cyclotron energies which are of order $E_c \sim 100 
{\rm K} \sim 10$ meV. 

All experiments are in principle described by a microscopic 
Hamiltonian describing the conduction electrons and their interactions,
but accurate calculations with this Hamiltonian are difficult to perform. 
Although it is not strictly necessary for our later discussion, it is 
useful to imagine this effective theory to be written \`a la Kivelson, Lee
and Zhang, as a system of interacting bosons, described by a field $\Phi$, 
coupled to a statistics field, $a_\mu$, with an odd statistics 
parameter, $\theta = (2n+1) \pi$. (As is well known, such a system 
is exactly equivalent in two dimensions to interacting fermions \cite{Zhang}.)

Because of the large hierarchy, $E_r \ll E_c$, one can imagine
integrating out the largely-irrelevant high-energy dynamics to derive
an effective lagrangian with which to describe the 
low-energy experiments. 
Although the direct, first-principles calculation of the low-energy effective
theory is usually as difficult as solving the full microscopic model, 
progress may be made inasmuch as the low-energy degrees of freedom
are more weakly interacting than are those at higher energies. In this
case it can be possible to extract precise predictions within controllable 
approximations, even when the same cannot be done with the full underlying
microphysical system \cite{Weinberg,Polchinski,Shankar}. When this is 
possible, direct appeal to the microscopic theory is only required to 
establish the validity of the assumed low-energy degrees of freedom, and 
need not play a crucial quantitative role in the comparison with experiment.

Our key assumption here -- and in \cite{PVD} -- is that these 
techniques may be applied to Quantum
Hall systems, where we will assume the low-energy dynamics can be 
described by a system of weakly-interacting charged quasi-particles or
vortices. The quasiparticles need not be electrons, which could well be 
strongly interacting in the microscopic theory, but 
are taken to be some effective description of the low-energy physics. 
For instance, motivated by the composite fermion picture \cite{Heinonen}, 
we take the quasiparticles to be fermions when describing the Laughlin 
plateaux (for which $\sigma_{xy} = 1/(2n+1)$ in our units, with $e^2/h=1$). 
On the other hand vortices will be assumed instead to govern 
the low-energy response of the Hall insulator. 

More concretely, for those phases described by quasiparticles, following
\cite{PVD} we use
the following effective Lagrangian, describing the low-energy/long-wavelength
interaction of a collection of {\it bosonic} charged quasi-particles, coupled
to electro-magnetic, $A_\mu$, and statistical gauge fields,
$a_\mu$, with statistical angle $\theta$:
\eq
\label{Lagr}
\Scl_\theta(\xi,a,A) = - \, {\pi \over 2 \theta} \; 
\epsilon^{\mu\lambda\nu} \, a_\mu \partial_\lambda a_\nu + 
\Scl_p(\xi,a+A).
\eeq
Here $\Scl_p(\xi,a+A)$ is the Lagrangian for the quasi-particles,
where $\xi_k$ is the position of the $k$-th particle,
\eq
\label{pact}
\Scl_p = \sum_k \left[\frac{m}{2} \dot{\xi}^\mu_k \dot{\xi}_{k\mu} 
-q \dot{\xi}^\mu_k (a+A)_\mu - V(\xi) \right] \delta[x-\xi_k(t)] , 
\eeq
with $m$ the quasi-particle mass, $q$ the charge and
$V(\xi)$ a potential representing other quasi-particle interactions with
their environment. Eq.~(\ref{pact}) represents the first few terms of
a derivative expansion of the low-energy quasiparticle lagrangian.

For the present applications it is important to keep in mind that the
electromagnetic field, $A_\mu$, which appears in this effective theory
is itself a low-energy effective field. It does not, in particular, 
include the large background magnetic field, $B$, whose presence the
quantum Hall effect requires. (Indeed, it cannot include such a large
field, since the motion within this field would involve energies of
order $\omega_c = eB/m$, which have been integrated out to obtain the
low-energy theory.) $A_\mu$ instead represents all of the weaker
fields in the low-energy part of the problem, including in particular
those fields which are applied in order to describe the system's
electromagnetic response. The dependence of low-energy quantities on 
on the background field $B$ is implicit in all of the parameters of
the effective theory, such as in the total number of particles or
vortices, the particle/vortex masses and couplings, {\it etc.}.

For those phases whose low-energy behaviour is described by vortices, we
instead use the general vortex action, which for our purposes has a very
convenient representation in terms of the vortex positions, $y_k$, and
a new gauge potential, $b_\mu$, which is a dual representation
of the scalar field which mediates the long-range interactions amongst
vortices \cite{Zhang}.
\eqa
\label{dualLagr}
\tilde\Scl_\theta(y,a,b,A) &=& - \, 
{\pi \over 2 \theta} \; \epsilon^{\mu\lambda\nu} \,
a_\mu \partial_\lambda a_\nu - \epsilon^{\mu\lambda\nu} \,
b_\mu \partial_\lambda (a_\nu + A_\nu) \nonumber\\
&& +\Scl_v(y,b),
\eeqa
Here $\Scl_v(y,b)$ is the Lagrangian for the vortex motion,
\eq
\label{vact}
\Scl_v = \sum_{\tilde k} \left[\frac{\tilde m}{2} \dot{y}^\mu_{\tilde k} \dot{y}_{{\tilde k}\mu} 
-\,{\tilde q} \; \dot{y}^\mu_{\tilde k} b_\mu -\widetilde V(y)
\right] \delta[x-y_{\tilde k}(t)], 
\eeq
with $\tilde m$ the vortex mass, $\tilde q$ the vortex charge (governing
its coupling to the field $b_\mu$) and $\widetilde V(y)$ represents 
possible vortex interaction terms.

The central property which we now assume of the quasiparticle and vortex 
effective lagrangians, and which underlies our subsequent conclusions, 
is that $\Scl_p(\xi,a)$ and $\Scl_v(y,b)$ have the same functional form
when considered as functionals of their respective arguments, $(\xi,a)$ or
$(y,b)$. A sufficient condition for this to be true -- at least at the lowest orders 
of the derivative expansion which suffice in the low-energy limit -- is when
interactions with the environment, $V(\xi)$ and $V(y)$, are negligible. (Although
sufficient, this condition might not be absolutely necessary.) 

It remains an intractable problem to solve even these effective
theories in any generality. However, if $\Scl_p(\xi,a)$ and $\Scl_v(y,b)$ 
do have the same functional form it is possible to relate the 
electromagnetic response for a system of
vortices to the response for a similar system of quasiparticles. It is
this relationship which we now derive, without making the assumption of linear
response in the fields $A_\mu$.

\section{The Experiments}

Before diving into the implications of particle-vortex similarity, it is
worth describing the evidence for particle-vortex duality beyond linear
response. Besides being an interesting topic in its own right, a 
description of these experimental results provides a sharper statement
of what it is that must be derived in the subsequent sections. 

\vtop{
\includegraphics{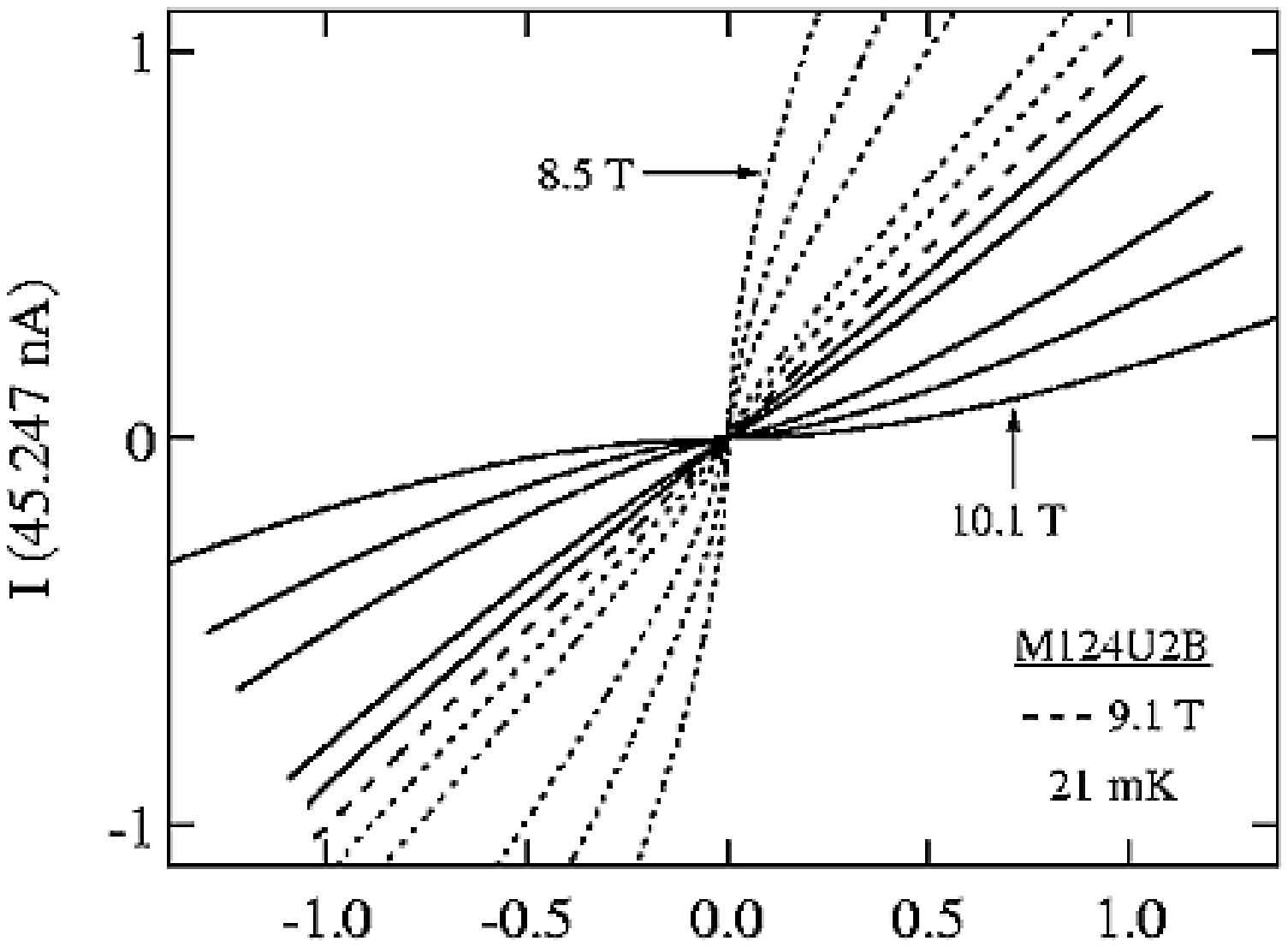}
\vskip 5.5cm
{\small Figure 1: Longitudinal current vs voltage in Quantum Hall
systems, taken for several magnetic fields on either side of the critical
point in the $\sigma_{xy} = 1$ to Hall insulator transition (reproduced
from ref.~\cite{nonlinear}). The solid lines correspond to one side of
the transition, and the dotted lines to the other side. Notice how the
solid lines are reflections of the dotted lines about the diagonal
line $I = V$, as would be expected for particle-vortex 
interchange. }}
\bigskip

The central nonlinear result is summarized by fig.~(1), which is reproduced
from ref.~\cite{nonlinear}. Each curve in this figure represents a trace
of the longitudinal (Ohmic) current, $I_x$, plotted against the
longitudinal voltage, $V_x$. The different curves are taken for different
values of the applied magnetic field, as the magnetic field is varied
across the transition between the $\sigma_{xy}=1$ quantum Hall plateau
and the Hall insulator. The solid lines are all traces taken on one side
of this transition, while the dotted lines are taken on the other side.

What is remarkable about this figure is its symmetry about reflection
through the diagonal axis $I_x = V_x$. Traces taken on one side of the 
transition are very accurately the reflection of those taken on the other
side of the transition. What is not shown in this figure, but is demonstrated 
in \cite{nonlinear}, is that the filling factor, $\nu$, corresponding to the
mirror-image traces are spaced an equal distance, $\Delta \nu = |\nu -
\nu_c|$, from the critical filling factor. 

The linear-response regime in these plots corresponds to the straight
segments near the origin, and within this regime the reflected current/voltage
curves have slopes which correspond to resistivities which are 
related by $\tilde\rho_{xx} = \rho_c^2/
\rho_{xx}$, where $\rho_c = \rho_{xx}(\nu_c) =
1$ is the universal value of the critical Ohmic resistivity 
for this transition. 

While $\rho_{xx}$ is varying as just described, what does $\rho_{xy}$ do?
The experiments show that $\rho_{xy}$ is a constant, and so 
does not vary at all through the transition \cite{nonlinear}.

In ref.~\cite{PVD} it was shown that if the particle and vortex effective
lagrangians have the same functional form, as assumed above, and if the
quasiparticles are fermions (or related to fermions in a precise way),
then this imposes a remarkable constraint
on the trajectory followed by a system in the conductivity plane as 
external parameters (like $B$) are varied. The constraint is that the 
trajectory must commute with a discrete group, $\Gamma_0(2)$, which may
be defined by the transformation $\sigma \to (a\sigma + b)/(2c \sigma + d)$,
where the integers $a,b,c$ and $d$ satisfy $ad - 2 bc = 1$. Here 
$\sigma$ denotes the convenient complex quantity 
$\sigma = \sigma_{xy} + i \sigma_{xx}$.

This symmetry provides an excellent description of fig.~(1) specialized
to the linear-response regime \cite{semicircle}. First, the symmetry 
predicts the critical resistivity in transitions between plateaux
to be universal, and to be given by $\rho_c=1$ for transitions to the
Hall insulator from the Laughlin sequence. Second, it also
implies for these transitions that $\rho_{xy}$ must be 
constant throughout the transition (called
the `semicircle law' because these trajectories are semicircles when drawn
in the $\sigma$ plane). Finally, there is a symmetry generator which maps
each semicircle trajectory onto itself, but with endpoints reversed, and
which corresponds precisely to particle-vortex interchange. As is shown
in \cite{semicircle}, this symmetry element is precisely equivalent to
the observed symmetry. 

For instance, for transitions between the $\nu =1$
plateau and the Hall Insulator (which are along the semicircle centred at
$\sigma = \frac12$ linking $\sigma = 1$ and $\sigma =  0$) this symmetry
acts in the following way:
\begin{equation}
\label{dualityaction}
\tilde{\sigma} = {\sigma - 1 \over 2 \sigma - 1} .
\end{equation}
Once restricted to the semicircular trajectories in the $\sigma$ plane 
-- which correspond to curves having constant $\rho_{xy}$ -- 
eq.~(\ref{dualityaction}) is precisely equivalent to the statement
$\tilde{\rho}_{xx} = 1/\rho_{xx}$. The analogous symmetry for other
transitions is obtained from this by acting with the group element
which takes (0,1) to the desired endpoints $(p_1/q_1,p_2/q_2)$. 
For example, for the $1/3 \to 0$ transition this gives the symmetry
generator $\tilde{\sigma} = (3\sigma -1)/(10\sigma -3)$, which
again corresponds to $\tilde{\rho}_{xx} = 1/\rho_{xx}$. 

To describe the data beyond linear response we must show that the curves
$I_x(V_x)$ get mapped into their inverses, $\tilde{V}_x(\tilde{I}_x)$, 
by the action of interchanging particles and vortices. This is most 
easily demonstrated by proving the equivalent statement for the 
tangents to these curves, which is:
\begin{equation}
\label{nonlinversion}
\tilde{\rho}_{xx}(\tilde{I}_x) = {d \tilde{V}_x \over d\tilde{I}_x}
= {dI_x \over dV_x} = {1 \over \rho_{xx}(I_x)} ,
\end{equation}
when evaluated along a trajectory for which $\rho_{xy}$ is constant.
Notice that the only difference between eq.~(\ref{nonlinversion}) 
and the corresponding result in linear response is the dependence
on $I_x$ and $\tilde{I}_x$ which is allowed in eq.~(\ref{nonlinversion})
but not in linear response.

\section{Particle-Vortex Duality in the Nonlinear Regime}

We now show how eq.~(\ref{nonlinversion}) follows from the similarity 
of the quasiparticle and vortex effective lagrangians. We do so in two
steps. The first step, already given in \cite{PVD}, is to derive an 
expression relating the quasiparticle nonlinear electromagnetic response for 
$\theta = \pi$ to the vortex nonlinear response for $\theta = -\pi$.
(Both $\theta=\pi$ and $\theta = -\pi$ -- or any other odd multiple of
$\pi$ -- are appropriate to transitions
from the Laughlin plateaux since the quasi-particles in this case
are fermions, such as in the composite fermion picture.) The second 
step is to derive an exact expression for how the electromagnetic 
response for vortices varies when $\theta$ is changed from $\pi$ to 
$-\pi$. 

\subsection{Step I}

We start with a system of fermionic quasiparticles described 
using $\theta=-\pi$ in \pref{Lagr}. This gives
\eqa
\label{pL}
\Scl_{\theta=-\pi}(\xi,a,A) &=& +\, {1\over 2} \epsilon^{\mu\nu\lambda}
a_\mu \partial_\nu a_\lambda + \Scl_{\rm kin}(\xi)\nonumber\\
&& + j^\mu(\xi) (a+A)_\mu - V(\xi).
\eeqa

On the other hand, the vortex system with $\theta=\pi$ in \pref{dualLagr} 
gives
\eqa
\label{vL}
\tilde\Scl_{\theta=\pi}(y,a,b,A) &=& -\, {1\over 2} \epsilon^{\mu\nu\lambda}
a_\mu \partial_\nu a_\lambda  -  \epsilon^{\mu\nu\lambda}
b_\mu \partial_\nu (a+A)_\lambda \nonumber\\
&& \qquad + \tilde\Scl_{\rm kin}(y) + 
\tilde{j}^\mu(y) b_\mu -\widetilde V(y).
\eeqa

These are used to generate the electromagnetic response functions, 
$\Gamma_\theta[A]$ and $\widetilde{\Gamma}_\theta[A]$, by evaluating the
following path integrals:
\eq
\label{EffAction}
e^{{i\over \hbar}\Gamma_\theta[A]} = \int [da_\mu(x)] \, \prod_k [d\xi^\mu_k(t)] \;
\exp\left[ {i\over \hbar}\int d^3x \Scl_\theta(\xi,a,A) \right]
\eeq
and 
\eqa
\label{EffAction2}
e^{{i\over\hbar}\tilde\Gamma_\theta[A]} 
&=& \int [da_\mu(x)] \, [db_\mu(x)] \\ 
&& \qquad \prod_{\tilde k} [dy^\mu_{\tilde k}(t)]
\exp\left[ {i \over \hbar}
\int d^3x \tilde{\cal L}_\theta (y,a,b,A)\right] \nonumber  .
\eeqa
Of course neither $\Gamma_{\theta=-\pi}[A]$ nor 
$\tilde\Gamma_{\theta=\pi}[A]$ 
can be calculated exactly, but the argument to follow implies a
relation between them that must always be true in the long wavelength
limit so long as the quasi-particles and vortices interactions at
low energy are similar (or negligible).

Our goal for these two systems is to use the similarity of the lagrangians
to relate the results of performing the path integrations.
To see this relation first shift 
$b_\mu\rightarrow b_\mu +A_\mu$ in \pref{EffAction2}, and then perform the 
Gaussian integral over $a_\mu$. The resulting effective lagrangian is
\eqa
\label{vL2}
\tilde\Scl_{\theta=\pi}(y,b,A) &=& +\, {1\over 2} \epsilon^{\mu\nu\lambda}
b_\mu \partial_\nu b_\lambda - \frac12 \, \epsilon^{\mu\nu\lambda}
A_\mu \partial_\nu A_\lambda \nonumber\\
&& \qquad  + \tilde\Scl_{\rm kin}(y) + \tilde{j}^\mu(y) 
(b+A)_\mu - \widetilde V(y) \nonumber\\
&:=&\tilde{\cal L}_{\theta=\pi}^\prime(y,b,A) - \frac12 \, 
\epsilon^{\mu\nu\lambda}
A_\mu \partial_\nu A_\lambda ,
\eeqa
where the second equation defines $\tilde{\cal L}_\theta$.

The main point is that  $\tilde{\cal L}_{\theta=\pi}$ has the same
form as does ${\cal L}_{\theta=-\pi}$, to the extent that both
$\tilde\Scl_{\rm kin}(y)$ and $\Scl_{\rm kin}(\xi)$
and $\tilde{j}^\mu(y)$ and $j^\mu(\xi)$ have the same functional
form. This ensures that $\widetilde{\Gamma}_{\theta=\pi}[A]$ is
related to $\Gamma_{\theta=-\pi}[A]$ by
\eq
\label{nonlindual}
\tilde\Gamma_{\theta=\pi}[A] = \Gamma_{\theta=-\pi}[A] - 
\frac12 \int d^3x \; \epsilon^{\mu\nu\lambda}
A_\mu \partial_\nu A_\lambda ,
\eeq
even though we cannot calculate either explicitly.
Notice that equation \pref{nonlindual} goes beyond linear response ---
there is no need to assume that $\Gamma_\theta[A]$ or 
$\tilde\Gamma_\theta[A]$ is quadratic in $A$. 

The relation between the nonlinear conductivities, (such
as $\sigma_{xx}(V_x)$) of the particle and
vortex systems is now obtained from by differentiating to
obtain the polarization tensor
\eq
\label{PiDef}
\Pi^{\mu\nu}_\theta=-{\delta^2 \Gamma_\theta[A] \over 
\delta A_\mu\delta A_\nu}.
\eeq
For a conductor the Fourier transformed quantity, 
$\Pi_\theta^{\mu\nu}(\omega,{\bf p})$, has a pole at 
$\omega=0$ and the conductivity is defined by
\eq
\label{sdef}
\sigma^\theta_{\alpha\beta}(A)=-i\lim_{\omega\rightarrow 0}
\left[\Pi^{\alpha\beta}_\theta(\omega,{\bf 0})/\omega\right],
\eeq
where $\alpha,\beta=x,y$. Notice that both of these definitions
also apply in the nonlinear regime, so long as eq.~\pref{PiDef}
is {\it not} evaluated at zero field: $A_\mu = 0$. 

For the dual system we see eq.~\pref{nonlindual} implies
\eq
\label{nlduallr}
\tilde\Pi^{\mu\nu}_{\theta=\pi}(p) = \Pi^{\mu\nu}_{\theta=-\pi}(p)
+ i\epsilon^{\mu\lambda\nu} p_\lambda,
\eeq
and so the nonlinear complex conductivities are related by
\eq
\label{T}
\tilde\sigma_{\theta=\pi}=\sigma_{\theta=-\pi} + 1.
\eeq
This is essentially the Landau level addition transformation of 
Kivelson, Lee and Zhang \cite{KLZ}, extended here to the non-linear regime.
The non-linear argument given here was first presented in \cite{PVD}.

\subsection{Step II: An Aside}

We next examine the effect on the electromagnetic response of a 
$2\pi$ shift of $\theta$ {\it without} interchanging particles with
vortices (or varying other external parameters). Once this is known,
it may be combined with eq.~\pref{T} to give the effect of 
particle-vortex interchange without simultaneously shifting $\theta$.

Before deriving the result of a $2\pi$ shift in $\theta$ we shall
pause to consider what it means. Indeed, one might reasonably expect
that {\it all} physical quantities -- and in particular the 
conductivities -- should be strictly periodic with respect to 
$\theta \to \theta + 2\pi$. To see why this need not be so in the
effective theory, we first review why it is true for the microscopic
theory. 

Within first-quantized theory the action describing the coupling
of the statistics field, $a_\mu$, to particles is strictly quadratic.
The path integral over $a_\mu$ is therefore Gaussian, and is equivalent
(up to an overall field-independent normalization) to evaluating the
action at its stationary point, $a_\mu = a^c_\mu$. Since this configuration
has vanishing field strength, $f^c_{\mu\nu} = 0$ (away from the position
of any of the particles to which it couples) it is locally pure gauge.
The integral $\oint a^c_\mu dx^\mu \ne 0$ about any curve which encloses
particle sources, however, so there is physics in $a^c_\mu$ and this
physics encodes the statistics phases which accrue whenever two particles
exchange positions \cite{ASW}.
 
If the particles involved all have hard cores and so can never interpenetrate
one another, then the particle positions may be excised and the 
physics of the statistics field comes purely from topology. In this case
the above picture gives the whole story, $a_\mu$ purely encodes particle
statistics, and all physical quantities are strictly periodic in $\theta$.
This is the situation for the microscopic electrons, such as described 
in the Quantum Hall context in ref.~\cite{Zhang}. 

The picture changes if the source for $a_\mu$ is distributed continuously.
Consider, for example, a uniform distribution of `charge' which gives
rise to a uniform distribution of statistical-field magnetic flux. In this
case the statistics field is not pure gauge since $f^c_{\mu\nu}\ne 0$, 
and its magnetic part is proportional to the source density. Consequently
physics can depend on the {\it local} values of $a_\mu$. Since
$a_\mu$ couples to $A_\mu$ only through the combination $(a+A)_\mu$,
particles see this magnetic statistics field as an addition
to the real magnetic field, $B$. 

For continously distributed source distributions, since $a_\mu$ 
encodes more than statistics phases there is no need for physical
quantities to be periodic under $\theta\to \theta + 2\pi$. Such aperiodicity
might be expected to occur in phases of the theory for which quasiparticles or 
vortices have condensed to form a nontrivial ground state.

For Quantum Hall systems we are led to a picture very much like that 
which arises in ref.~\cite{KLZ}.  For the microscopic electrons
the physics is strictly periodic under $2\pi$ 
shifts of the statistics angle, $\theta$.
However, the system has a great many phases, and the effective theory
built over the ground state of any particular phase need not 
be invariant under these shifts of the statistics angle. The periodicity
of the full theory is seen once all of these phases are viewed together,
since changes to $\theta$ take one phase into another. The change
of phase can be understood qualitatively because changes in $\theta$ cause
changes to $a^c_\mu$ which may be compensated by changes in $A_\mu$, and in 
particular to the applied magnetic field. But changing the applied magnetic
field is one of the methods used to move between different phases in
the lab.

Thus, the underlying invariance with respect to 
$\theta \to \theta + 2\pi$ emerges in the effective theory
as a relation between the properties of {\it different} phases of 
the system, with the physics of any individual phase not being
simply periodic. We may legitimately ask what the action of
such a shift is on the electromagnetic response of the system.

\subsection{Step II: The Calculation}

We now proceed with the calculation of the effects of a $2\pi$
shift of $\theta$ on the nonlinear response function,
$\Gamma_\theta[A]$. To this end consider the generating 
function, $W_\theta[J]$, for the
electromagnetic correlation functions:
\eq
\exp\left({i\over \hbar} W_\theta[J]\right) = 
\int [dA] \exp\left({i\over \hbar}\Gamma_\theta[A] 
+ {i\over \hbar} \int d^3x A_\mu J^\mu\right),
\eeq
where $\Gamma_\theta[A]$ is defined by (\ref{Lagr}) and (\ref{EffAction}).
To perform the $a$ integral shift $A_\mu \to B_\mu := A_\mu
+ a_\mu$, so the statistics field $a_\mu$ only appears in the Chern-Simons term
in (\ref{pL}) and
through the current coupling $\int d^3x (B_\mu - a_\mu)J^\mu$. The 
$a_\mu$ integral may then be explicitly performed, 
since it is Gaussian. The result is (neglecting
as usual overall factors):
\eqa
&&\exp\left({i\over \hbar} W_\theta[J]\right)=\int [dB]
\exp\left\{ {i\over \hbar} S[B] +{i\over \hbar}
\int d^3x B_\mu J^\mu \vline height15pt width 0pt \right. \nonumber\\
&&\left. - {i\over \hbar}\left({\theta \over 2 \pi}
\right) \int\int d^3x d^3x' \epsilon^{\mu\nu\lambda} 
J_\mu \left({1\over \partial^2} \right)
\partial_\nu J_\lambda \right\},\nonumber\\
\eeqa
where $e^{{i\over\hbar} S[B]}=\int \Pi_k[d\xi_k]e^{{i\over\hbar}
\int d^3x {\cal L}_p(\xi,B)}$

This makes the $\theta$-dependence of $W_\theta[J]$ explicit, 
so 
\eq W_\theta[J] = W_0[J] - \left({\theta \over 2 \pi}
\right) \int\int d^3x\, d^3x' \epsilon^{\mu\nu\lambda} J_\mu 
\left({1\over \partial^2} \right)\partial_\nu J_\lambda' .
\eeq

To make contact with the polarization tensor, $\Pi_\theta^{\mu\nu}$,
we must relate $W_\theta[J]$ to $\Gamma_\theta[A]$. To within a very
good approximation they are Legendre transforms of one another. That is,
defining the Legendre transform, $L_\theta[A]$, of $W_\theta(J)$ by:
\eq
L_\theta[A] = W[J] - \int d^3x A_\mu J^\mu,
\eeq
with $A_\mu = \delta W_\theta /\delta J^\mu$, 
standard field-theoretic arguments imply that $L_\theta$ is related
to $\Gamma_\theta$ in the following way:
\eq
e^{{i\over \hbar} L_\theta[A]} = \int [dA'] e^{{i\over \hbar} \left[
\Gamma_\theta[A'+A] + \int A'_\mu J^\mu[A] dx \right]}
\eeq
where $J^\mu = - \delta L_\theta/\delta A_\mu$. It follows 
that $\Gamma_\theta[A]$ 
and $L_\theta$ are equal to one another if the $A'_\mu$ integral is performed
semiclassically. Since the low-energy
applied electromagnetic fields used in linear response, $A_\mu$, are very 
well described semiclassically, we can equate $L_\theta$ and $\Gamma_\theta$
to equally good approximation.

It then follows that the derivative $W^\theta_{\mu\nu} = \delta^2W_\theta
/\delta J^\mu \delta J^\nu$ is related to $\Pi_\theta^{\mu\nu}$ by
$\Pi_\theta^{\mu\nu} W^\theta_{\nu \lambda} = \Lambda_\lambda^\mu$, where
$\Lambda_{\mu\nu} = \eta_{\mu\nu} - p_\mu p_\nu/p^2$. We use here 
(for convenience of notation only) a relativistic
notation with $\eta_{\mu\nu}=\hbox{diag}(-1,1,1,1)$ \cite{PVD}. For
brevity we write this relation as $W^\theta_{\mu\nu} =
(\Pi_\theta^{\mu\nu})^{-1}$.

Combining the above results, in momentum space we have:
\eq
\left(\Pi^{\mu\nu}_\theta\right)^{-1} \approx 
\left(\Pi^{\mu\nu}_0\right)^{-1} + \left( {\theta \over \pi} \right)
\; {\hbar\over \sqrt{p^2}} \; {\cal J}^{\mu\nu}, 
\eeq
where ${\cal J}_{\mu\nu} = i \epsilon_{\mu\lambda\nu} p^\lambda/\sqrt{p^2}$.

For $\theta=2\pi$ this reproduces the results of \cite{PVD}
for the flux attachment transformation for the conductivities,
\eq
\label{ST2S}
-{1\over\tilde\sigma}=-{1\over\sigma} +2,
\eeq
where $\tilde\sigma$ is obtained from $\Pi^{\mu\nu}_\theta$ as in
\pref{sdef}. This is the flux attachment transformation of \cite{KLZ},
extended again to the non-linear regime.  The only difference
between eq.~\pref{ST2S} and the linear-regime results of \cite{PVD}
is that here $\sigma$ can be a function of the external 
electro-magnetic effective field.

\subsection{Particle-Vortex Interchange}

Our goal is to derive eq.~\pref{nonlinversion} as 
the effect of particle-vortex interchange (at fixed $\theta$),
and so we must combine the results of eqs.~\pref{T} and \pref{ST2S}.

The simplest way to do so is to recognize the 
group -- $\Gamma_0(2)$ -- which is obtained
through repeated applications of equations \pref{T} and 
\pref{ST2S} \cite{LutkenRoss,KLZ,Lutken}.
A familiar form for this group structure is most easily seen 
by writing it in terms of the two operations
\eqa
T:\sigma&\rightarrow& \sigma +1\nonumber\\
S:\sigma&\rightarrow& -{1\over \sigma},
\eeqa
which satisfy $(ST)^3=1$. In terms of these operations
the group of interest ($\Gamma_0(2)$) is generated by
\eqa
\label{G02}
T:\sigma&\rightarrow& \sigma +1\nonumber\\
ST^2S:\sigma&\rightarrow& {\sigma \over 1-2\sigma},
\eeqa

The operation $S$ is only introduced
here for convenience. It is not a symmetry
of the quantum Hall effect, since it cannot be obtained by
repeated applications of the basic transformations, \pref{T}
and \pref{ST2S}. ($S$ represents interchange of the
conductivity and the resistivity.  Although it is not a symmetry of the
Quantum Hall Effect, for which the charge carriers are fermions, it 
should be a symmetry for 2-dimensional systems in which the charge carriers
are bosonic, \cite{PVD}.)

Combining the two operations $T$ and $ST^2S$ we see that
the effect of interchanging fermionic quasiparticles and vortices 
is given by 
\eq
\label{TST2S}
T ST^2S(\sigma)=  {\sigma-1\over 2\sigma -1} ,
\eeq
which is precisely eq.~\pref{dualityaction}, although now generalized
to the nonlinear regime by including field-dependent $\sigma$.

As was discussed earlier, this provides a successful description of
the nonlinear duality of the transition between the $\sigma_{xy}= 1$
plateau and the Hall insulator. Since the group structure is the
same as in the linear-response regime, we may now repeat the 
linear-response arguments \cite{semicircle} to immediately 
understand the analogous result for particle-vortex interchange
in the transitions to the Hall insulator from the Laughlin sequence,
$\sigma_{xy} = 1/(2n+1)$, despite the fact that the quasiparticles
in this instance enjoy fractional statistics, and so are no longer
fermions. 

To understand the $\nu={1\over 3}\rightarrow 0$ transition in the
language presented here, we obtain it as a symmetry transformation of
the  $\nu=1\rightarrow 0$ transition just described.  The modular 
symmetry \pref{G02} dictates that the $1 \to 0$ transition follows a 
semi-circle in the complex $\sigma$-plane, with the critical point
at $\sigma_c=(1+i)/2$, \cite{semicircle}, and we have found
quasi-particle -- vortex duality to be implemented by the 
transformation, eq.~\pref{TST2S}.

This transformation interchanges the end points 
$0\leftrightarrow 1$ and leaves the critical point, $\sigma_c$, 
fixed. To study other transitions, such as the $\nu={1\over 3}\rightarrow 0$ 
transition examined experimentally,
we must find the group element which maps this basic 
semi-circle of radius ${1\over 2}$
arching between $\sigma=1$ and $\sigma=0$, onto the semi-circle
of radius ${1\over 6}$ arching between $\sigma={1\over 3}$ and $\sigma=0$.
Once found, this group element can be used to transform 
eq.~\pref{dualityaction} to the transformation appropriate for 
particle-vortex interchange in the $\frac13 \to 0$ transition.
The result of this exercise is the transformation $\sigma \to
(3\sigma -1)/(10 \sigma - 3)$. As is easily checked, this
interchanges $\sigma=\frac13$ and $\sigma=0$, and 
maps the critical point, $\sigma_c={3+i\over 10}$, to itself.

Once mapped to the resistivity plane, the transition is again
along the line with constant $\rho_{xy} = 3$, along which 
particle-vortex interchange becomes  $\rho_{xx} \to 1/\rho_{xx}$.
This shows that the experimental observations of \cite{nonlinear} 
are the consequence of the particle-vortex interchange,
even deep within the non-linear regime. Just as for linear response \cite{PVD}, 
this effective field theory analysis sheds light
on why the duality is experimentally successful so far
from the critical points, to which the analysis of \cite{KLZ} was
believed to be restricted. 

In conclusion we have shown that the Law of Corresponding States
is applicable in the non-linear regime --- well outside of the linear
regime of its original derivation. In particular the pseudo-particle -- vortex
duality which was invoked to explain the experiments in \cite{nonlinear}
can be extended into the non-linear regime, as is necessary to explain
these experimental observations.

Our arguments assume the long-wavelength, low-energy limit
and are applicable to any system in which the interactions between the pseudo-particles
are weak, as are the interactions between the vortices, so that there is a symmetry
under interchange of pseudo-particles and vortices.  More generally one could allow
stronger interactions between the pseudo-particles provided the interaction
potential between vortices are of the same form, but this might be harder to
realise in practice. 

For fermionic pseudo-particles the resulting symmetry
group is $\Gamma_0(2)$ and this group therefore
seems to be the one relevant to the quantum Hall 
effect (Other candidates have also been considered, \cite{G2},
which may be the relevant symmetries when electron spins are not
well separated by Zeeman splitting
\cite{espin}).  
It was argued in \cite{PVD} that a different group is relevant when
the pseudo-particles are bosonic --- the group given by matrices
$\gamma=\left(\matrix{ a&b\cr c& d\cr}\right)$ with $ad$ even and $bc$ odd or
vice versa. This group is often denoted by $\Gamma_\theta(2)$ in the
mathematical literature and is generated by $S$ and $T^2$.  The arguments
presented here are, of course, just as applicable to such bosonic systems.

We thank A. MacDonald and J. Maldacena for helpful discussions. 
C.B. is grateful to Aspen Centre for Physics where part of this work was carried
out. Our research has been assisted by financial support from N.S.E.R.C. (Canada), 
F.C.A.R. (Qu\'ebec), the Ambrose Monell Foundation and Enterprise 
Ireland, Basic Research Grant no. SC/1998/739.


\begin{thebibliography}{99}
\bibitem{Caveat}
The experimental situation more recently has become more murky, 
with new evidence for \cite{ForScaling} and against \cite{AgainstScaling}
scaling behaviour. We predict scaling and universality to occur in any
system to which the particle-vortex duality we propose applies.
\end{thebibliography}
\end{document}